\documentstyle[multicol,aps,prl,epsfig,graphicx,array]{revtex}

\renewcommand{\narrowtext}{\begin{multicols}{2} \global\columnwidth20.5pc}
\renewcommand{\widetext}{\end{multicols} \global\columnwidth42.5pc}
\def\top#1{\vskip #1\begin{picture}(290,80)(80,500)\thinlines \put(
65,500){\line( 1, 0){255}}\put(320,500){\line( 0, 1)
{ 5}}\end{picture}}
\def\bottom#1{\vskip #1\begin{picture}(290,80)(80,500)\thinlines \put(
330,500){\line( 1, 0){255}}\put(330,500){\line( 0, -1){
5}}\end{picture}}
\def\wtxt{\widetext \top{-2.8cm} \hglue -1 cm}
\def\ntxt{ \bottom{-2.7cm} \narrowtext \noindent}

\def\al{\alpha}
\def\be{\beta}
\def\ga{\gamma}

\def\ep{\epsilon}
\def\ve{\varepsilon}

\def\et{\eta}

\def\ka{\kappa}
\def\la{\lambda}

\def\rh{\rho}

\def\si{\sigma}

\def\ps{\psi}

\def\Ga{\Gamma}

\def\fr#1#2{{{#1} \over {#2}}}
\def\frac#1#2{\textstyle{{{#1} \over {#2}}}}
\def\half{{\textstyle{1\over 2}}}
\def\pt#1{\phantom{#1}}
\def\prt{\partial}

\def\lsim{\mathrel{\rlap{\lower4pt\hbox{\hskip1pt$\sim$}}
    \raise1pt\hbox{$<$}}}
\def\gsim{\mathrel{\rlap{\lower4pt\hbox{\hskip1pt$\sim$}}
    \raise1pt\hbox{$>$}}}

\newcommand{\bequ}{\begin{equation}}
\newcommand{\eequ}{\end{equation}}
\newcommand{\beq}{\begin{eqnarray}}
\newcommand{\eeq}{\end{eqnarray}}
\newcommand{\bea}{\begin{eqnarray}}
\newcommand{\eea}{\end{eqnarray}}
\newcommand{\rf}[1]{(\ref{#1})}

\def\kkaf{{$(k_{AF})_\mu$}}
\def\kkf{{$(k_{F})_{\ka\la\mu\nu}$}}

\def\a{$a_\mu$}
\def\b{$b_\mu$}
\def\c{$c_{\mu\nu}$}
\def\d{$d_{\mu\nu}$}
\def\e{$e_\mu$}
\def\f{$f_\mu$}
\def\g{$g_{\la\mu\nu}$}
\def\H{$H_{\mu\nu}$}

\def\etal{{\it et al.}}

\newcommand{\fslash}[1]{#1 \!\!\! /}
\newcommand{\incpicwh}[3]{\includegraphics[width=#2,height=#3]{#1}}
\newcommand{\pad}{\hspace{-0.5cm}}
\newcommand{\fgl}[1]{\hspace{1.05cm}#1\hspace{-1.05cm}}
\newcommand{\qedpic}[1]{\incpicwh{#1}{2.6cm}{2.6cm}}

\begin{document}

\title{Vacuum Photon Splitting in 
       Lorentz-Violating Quantum Electrodynamics}
\author{V.\ Alan Kosteleck\'y and Austin G.M.\ Pickering}
\address{Physics Department, Indiana University, 
       Bloomington, IN 47405}
\date{IUHET 453, December 2002; 
accepted for publication in Physical Review Letters} 
\maketitle

\begin{abstract}
Radiative corrections arising from Lorentz violation
in the fermion sector induce a nonzero amplitude 
for vacuum photon splitting.
At one loop, the on-shell amplitude 
acquires both CPT-even and CPT-odd contributions
forbidden in conventional electrodynamics.

\end{abstract}


\narrowtext

In quantum electrodynamics (QED),
a photon traveling in the vacuum has a zero amplitude for 
decay into multiple on-shell photons
at any order in perturbation theory.
This classic no-splitting result is a byproduct 
of Schwinger's work on the proper-time method
and the Euler-Heisenberg effective action in QED
\cite{js}.

Schwinger's result relies heavily
on gauge invariance and also implicitly 
on Lorentz and CPT symmetry.
However,
Lorentz and CPT symmetry may be broken by effects 
from the Planck scale
\cite{kps}.
Any effects are expected to be heavily suppressed
by at least one power of the Planck mass,
which implies most processes in QED and the Standard Model
acquire only small Lorentz-violating corrections.
Nonetheless,
some processes normally excluded may now occur.
It is natural to ask how Schwinger's result 
forbidding on-shell vacuum photon splitting
generalizes to the Lorentz-violating case. 

The purpose of this work is to 
discuss the amplitude for vacuum photon splitting
in the context of the general Lorentz- and CPT-violating QED extension
\cite{ck},
which includes all possible Lorentz-violating terms involving 
electron and photon fields.
At leading order 
in certain coefficients for Lorentz violation,
we find a nonzero on-shell one-loop amplitude 
for photon triple splitting.

We work within the renormalizable sector
of the general QED extension
for a single Dirac field $\ps$ of mass $m$,
for which the lagrangian $\cal L$ is
\cite{ck,nc} 
\bea
{\cal L} &=& \frac{1}{2} i \bar{\ps} \Ga^\nu
\stackrel{\leftrightarrow}{D_\nu} \ps  -  \bar{\ps}
M \ps  - \frac{1}{4} F^{\mu\nu} F_{\mu\nu} 
\nonumber \\ & & 
- \frac {1}{4}
(k_F)_{\ka\la\mu\nu} F^{\ka\la} F^{\mu\nu} + \frac{1}{2}
(k_{AF})^{\ka} \ep_{\ka\la\mu\nu} A^\la F^{\mu\nu},
\label{lag}
\eea
where $D_\mu$ is the usual covariant derivative and
\beq
\Ga^{\nu} &\equiv&
\ga^{\nu} + c^{\mu\nu} \ga_\mu + d^{\mu\nu} \ga_5
\ga_\mu + e^\nu + i f^\nu \ga_5 + \half g^{\la\mu\nu}
\si_{\la\mu},
\nonumber \\
M &\equiv& m + a_\mu \ga^\mu + b_\mu \ga_5 \ga^\mu +
\half H_{\mu\nu} \si^{\mu\nu}.
\label{coeff}
\eeq
The Lorentz violation is controlled by the real coefficients
\a, \b, \c, \d, \e, \f, \g, \H\ in the fermion sector
and \kkaf, \kkf\ in the photon sector.
Inspection of Eq.\ \rf{lag} shows the amplitude
for vacuum photon splitting is zero at tree level.
The task at hand is thus to investigate possible 
finite radiative corrections to photon splitting at one loop
and at leading order in the coefficients for Lorentz violation.
In fact,
for our purposes it suffices to restrict attention to 
a conventional photon sector with negligible \kkaf, \kkf.
The four coefficients \kkaf\ have been strongly constrained using 
measurements of cosmological birefringence
\cite{cfj,jk,adkl}.
Ten of the 19 independent coefficients in \kkf\
have also been strongly constrained 
using spectropolarimetry of cosmological sources,
while the other nine can be absorbed into the 
fermion sector by a field redefinition without loss of generality 
\cite{km}.

With a conventional photon sector,
photon splitting in the vacuum is strongly restricted by kinematics. 
If the incident on-shell photon has 
energy $E_1$ and 3-momentum $\vec p_1$, 
and the $n$ photons produced in the decay have 
energies $E_j$ and 3-momenta $\vec p_j$, 
$j = 2,3,\ldots, n+1$,
then conservation of 3-momentum implies 
$\sum_j |\vec p_j| \geq |\vec p_1|$.
Since each on-shell photon has 4-momentum $p^\mu = (E, \vec p)$
satisfying $p^2 = 0$ and hence $E = |\vec p|$,
this inequality is compatible with conservation of energy 
only if all the $\vec p_j$ are aligned. 
The incident photon and the decay products 
must therefore be collinear.
It then follows that the 4-momenta of all photons
are mutually orthogonal,
$p_j^\mu p_{k\mu} = 0$,
and that they are all proportional to some momentum $p_0^\mu$ 
satisfying $p_0^2 = 0$ on shell.
Moreover,
the transversality of each physical photon
implies that its polarization 4-vector $\ve^\mu$
obeys $p_\mu \ve^\mu = 0$.
Together with the requirement of collinearity,
this implies $\ve_j^\mu p_{k\mu} = 0$
and that there are at most two linearly independent 
physical polarization vectors in any process. 

In addition to these kinematical constraints,
the amplitude for photon splitting must satisfy criteria
imposed by symmetry transformations and field redefinitions.
First,
since the theory \rf{lag} is invariant under observer Lorentz 
transformations,
the amplitude must be an observer Lorentz scalar
and so any Lorentz indices it contains must be contracted.
Second,
the properties of the Lorentz-violating operators in Eq.\ \rf{lag} 
under charge conjugation C
imply a generalization of Furry's theorem,
which eliminates both divergent and finite contributions 
arising from certain coefficients 
\cite{klp}.
It follows that
only \a, \d, \e, \f, \H\ can contribute
to splitting into an even number of photons, 
while only \b, \c, \g\ can contribute 
to odd splitting.
Third,
the behavior of Eq.\ \rf{lag} under parity inversion P ensures
that any contribution to the amplitude linear in 
\b, \d, \f\ must come with one factor of the antisymmetric tensor 
$\ep_{\mu\nu\rh\si}$. 
Fourth, 
gauge invariance demands that 
the amplitude be invariant when 
an external photon polarization vector is shifted 
by an amount proportional to its momentum.
The expression for the amplitude must therefore vanish 
if any one polarization vector is replaced 
by the corresponding external momentum.
Finally,
in the single-fermion theory \rf{lag},
some coefficients for Lorentz violation can be removed 
at linear order by suitable field redefinitions 
\cite{ck}
and cannot contribute at leading order to physical processes.
So we can ignore potential contributions 
from \a, \e, \f, 
from the antisymmetric parts of \c, \d, 
and from all but the mixed-symmetry part of \g. 

The combination of all the above constraints
severely restricts the possibilities for photon splitting.
For splitting into two photons,
we find the only possibility for a nonzero amplitude
involves the combination
$\ve_j^\mu H_{\mu\nu} p_k^\nu \ve_l^\al \ve_{m\al}$.
For splitting into three photons,
the possibilities for a nonzero amplitude
include the two combinations
$\ep_{\mu\nu\rh\si} b^\mu p^\nu 
\ve_j^\rh \ve_k^\si \ve_l^\al \ve_{m\al}$
and 
$c_{\mu\nu} p^\mu p^\nu \ve_j^\al \ve_{k\al} \ve_l^\be \ve_{m\be}$.
Similar strong constraints can be deduced 
on the amplitudes for splitting into any number of photons.

To determine whether any of these amplitudes are nonzero,
we proceed by direct calculation. 
For splitting into two photons,
the calculation reveals that the overall amplitude vanishes. 
Photon double splitting is therefore absent,
despite the presence of general Lorentz-violating terms 
in the fermion sector.
However,
the calculation of photon triple splitting
reveals a nonzero amplitude,
as we demonstrate next
\cite{fn1}. 

\begin{figure}
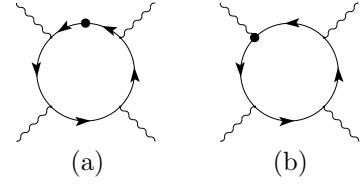

\centering
\begin{tabular}{cc}
\fgl{(a)}\pad\qedpic{p4prop11.eps}\pad & 
\fgl{(b)}\pad\qedpic{p4vert11.eps}\pad \\ ~
\end{tabular}
\caption{Diagrams for one-loop photon triple splitting.}
\label{insertgphs}
\end{figure}

Consider first the P- and CPT-even contribution 
to photon triple splitting,
which is linear in \c. 
This arises from a propagator insertion involving \c\ 
in the usual one-loop four-point diagram,
as shown in Fig.\ 1a.
Note that a vertex insertion of \c, 
shown in Fig.\ 1b, 
cannot give a gauge-invariant contribution 
to on-shell photon triple splitting 
because \c\ always appears contracted with a polarization vector.
Denoting the external momenta as $p_j$, 
$j = 1,2,3,4$, and the loop momentum as $k$,
and adopting the shorthand notation
$k_1 = k + p_1$, $k_{12} = k + p_1 + p_2$, 
etc.,
the contribution $G$ from 
a single graph of the form in Fig.\ 1a is
\wtxt
\beq
G^{\mu_1 \mu_2 \mu_3 \mu_4} 
&=& \int \fr{d^4k}{(2\pi)^4} \left[
\fr{
{\rm Tr} [ 
(\fslash{k}+m) c^{\mu\nu} \ga_\mu k_\nu 
(\fslash{k}+m) \ga^{\mu_1}
(\fslash{k}_{1}+m)\ga^{\mu_2}
(\fslash {k}_{12}+m)\ga^{\mu_3} 
(\fslash{k}_{123}+m)\ga^{\mu_4}
]}
{(k^2-m^2)^2(k_1^2-m^2)(k_{12}^2-m^2)(k_{123}^2-m^2)} 
\right].
\eeq
The denominators can be combined 
in the usual way using Feynman parameters,
so that $G = A + B$ with
\beq
A &=& 4! 
\int_0^1 dx_1 
\int_0^{1-x_1} \hspace{-2em} dx_2 
\int_0^{1-x_{12}} \hspace{-2em} dx_3 
\int_0^{1-x_{123}} \hspace{-2em} dx_4 
\int \fr{d^4k}{(2\pi)^4}~ 
2 k_\mu c^{\mu\nu} k_\nu 
\left[
\fr{
{\rm Tr} 
[(\fslash{k}+m)\ga^{\mu_1}
(\fslash{k}_{1}+m)\ga^{\mu_2}
(\fslash{k}_{12}+m)\ga^{\mu_3} 
(\fslash{k}_{123}+m)\ga^{\mu_4}
]}
{(k^2-m^2 + 2 k \cdot q)^5} 
\right],
\nonumber \\
B &=& - 3! 
\int_0^1 dx_1 
\int_0^{1-x_1} \hspace{-2em} dx_2 
\int_0^{1-x_{12}} \hspace{-2em} dx_3 
\int \!\!\! \fr{d^4k}{(2\pi)^4} 
\left[
\fr{
{\rm Tr} 
[(c^{\mu\nu} \ga_\mu k_\nu)\ga^{\mu_1}
(\fslash{k}_{1}+m) \ga^{\mu_2}
(\fslash{k}_{12}+m)\ga^{\mu_3} 
(\fslash{k}_{123}+m) \ga^{\mu_4}
]}
{(k^2-m^2 + 2 k \cdot q)^4} 
\right],
\eeq
where 
$x_{12}=x_1+x_2$, $x_{123}=x_1+x_2+x_3$,
$q = x_1 p_1 + x_2 p_{12} + x_3 p_{123}$ with 
$q^2 = 0$, 
and the indices on $A$, $B$ are suppressed for brevity. 
The trace of an odd number of $\ga$-matrices vanishes, 
so the traces can be expanded as polynomials in $m^2$: 
\beq
A &=& 4! 
\int_0^1 dx_1 
\int_0^{1-x_1} \hspace{-2em} dx_2 
\int_0^{1-x_{12}} \hspace{-2em} dx_3 
\int_0^{1-x_{123}} \hspace{-2em} dx_4 
( A_1 + m^2 A_2 + m^4 A_3 ) ,
\quad 
B = 3! 
\int_0^1 dx_1 
\int_0^{1-x_1} \hspace{-2em} dx_2 
\int_0^{1-x_{12}} \hspace{-2em} dx_3 
(B_1 + m^2 B_2 ) ,
\label{AB}
\eeq
where
\beq
A_1 &=& 
\int \fr{d^4k}{(2\pi)^4} ~2 k_\mu c^{\mu\nu} k_\nu 
\left[
\fr{
{\rm Tr} [
\fslash{k}\ga^{\mu_1}
\fslash{k}_{1}\ga^{\mu_2}
\fslash{k}_{12}\ga^{\mu_3} 
\fslash{k}_{123}\ga^{\mu_4}
]}
{(k^2-m^2 + 2 k \cdot q)^5} 
\right],
\nonumber \\ 
A_2 &=& 
\int \fr{d^4k}{(2\pi)^4} ~ 2 k_\mu c^{\mu\nu} k_\nu 
\Big[
\fr{
{\rm Tr} [
\fslash{k}\ga^{\mu_1}
\fslash{k}_{1}\ga^{\mu_2} \ga^{\mu_3}\ga^{\mu_4}
] 
+ {\rm Tr} [
\fslash{k}\ga^{\mu_1}\ga^{\mu_2}
\fslash{k}_{12}\ga^{\mu_3} \ga^{\mu_4}
] 
+ {\rm Tr} [
\fslash{k}\ga^{\mu_1} \ga^{\mu_2} \ga^{\mu_3} 
\fslash{k}_{123} \ga^{\mu_4}
] }
{(k^2-m^2 + 2 k \cdot q)^5} 
\nonumber \\ &&
\hskip 90pt
+ \fr{
{\rm Tr} [
\ga^{\mu_1}
\fslash{k}_{1}\ga^{\mu_2}
\fslash{k}_{12}\ga^{\mu_3} \ga^{\mu_4}
] 
+ {\rm Tr} [
\ga^{\mu_1}
\fslash{k}_{1}\ga^{\mu_2}
\ga^{\mu_3} \fslash{k}_{123}\ga^{\mu_4}
] 
+ {\rm Tr} [
\ga^{\mu_1}\ga^{\mu_2}
\fslash{k}_{12}\ga^{\mu_3} 
\fslash{k}_{123}\ga^{\mu_4}
] }
{(k^2-m^2 + 2 k \cdot q)^5} 
\Big],
\nonumber \\
A_3 &=& \int \fr{d^4k}{(2\pi)^4} ~ 2 k_\mu c^{\mu\nu} k_\nu 
\left[
\fr{ 
{\rm Tr} [
\ga^{\mu_1}\ga^{\mu_2}\ga^{\mu_3}\ga^{\mu_4}] 
}
{(k^2-m^2 + 2 k \cdot q)^5} 
\right],
\eeq
and
\beq
B_1 &=& - \int \hspace{-2pt} \fr{d^4k}{(2\pi)^4} 
\left[
\fr{
{\rm Tr} [
\ga^{\mu} c_{\mu\nu} k^{\nu} \ga^{\mu_1}
\fslash{k}_{1}\ga^{\mu_2}
\fslash{k}_{12}\ga^{\mu_3} 
\fslash{k}_{123}\ga^{\mu_4}]
}
{(k^2-m^2 + 2 k \cdot q)^4} 
\right],
\nonumber \\ 
B_2 &=& - \int \hspace{-2pt} \fr{d^4k}{(2\pi)^4} 
\left[
\fr{
{\rm Tr} [
\ga^{\mu} c_{\mu\nu} k^{\nu} \ga^{\mu_1}
\fslash{k}_{1}\ga^{\mu_2} \ga^{\mu_3}\ga^{\mu_4}
] 
+ {\rm Tr} [
\ga^{\mu} c_{\mu\nu} k^{\nu} \ga^{\mu_1}\ga^{\mu_2}
\fslash{k}_{12}\ga^{\mu_3} \ga^{\mu_4}
] 
+ {\rm Tr} [
\ga^{\mu} c_{\mu\nu} k^{\nu} \ga^{\mu_1}\ga^{\mu_2} \ga^{\mu_3} 
\fslash{k}_{123}\ga^{\mu_4}
] }
{(k^2-m^2 + 2 k \cdot q)^4} 
\right] .
\eeq
The traces in these expressions can be reduced using
\beq
2 {\rm Tr} [\ga_1\ga_2\ldots\ga_n ] &=& 
\et_{12} {\rm Tr} [ \ga_3\ga_4\ldots\ga_n]
- \et_{13} {\rm Tr} [\ga_2\ga_4\ldots\ga_n] 
+ \et_{14} {\rm Tr} [\ga_2\ga_3\ldots\ga_n] 
- \ldots 
+ \et_{1n} {\rm Tr} [\ga_2\ga_3\ldots\ga_{n-1}] .
\label{traces}
\eeq
The full contribution to this vertex 
from all the relevant graphs,
including all permutations over external momenta,
must be finite
\cite{klp}.
However,
the expression $G$ for a single contributing graph diverges,
and the divergences must be isolated by regulation.
The integrals can be performed in $d = 4 - 2\ep$ dimensions,
using the result 
\beq
\int \fr{d^dk}{(2\pi)^4} 
\fr{k_{\mu_1}\ldots k_{\mu_n}}
{(k^2-m^2 + 2 k \cdot q)^\alpha}
&=& 
(-1)^n 
\fr{i \pi^{d/2} (-m^2)^{(d-2 \al)/2}}{\Ga{(\al)}} 
\Big[ \Ga(\al-d/2) q_{\mu_1}\ldots q_{\mu_n} 
\nonumber \\ &&
\hskip 70pt
+ (-m^2/2) \Ga(\al - 1 - d/2)
( q_{\mu_1}\ldots q_{\mu_{(n-2)}} \et_{\mu_{(n-1)}\mu_n} 
+ {\rm permutations}) 
\nonumber \\ &&
\hskip 70pt
+ \ldots 
+ (-m^2/2)^{n/2} \Ga(\al - (n+d)/2)
\et_{\mu_1\mu_2} \ldots  \et_{\mu_{(n-1)}\mu_n}
\Big].
\eeq

Since the external momenta are parallel, 
they can all be written in terms of a single vector 
$p_0^\mu$ satisfying $p_0^2 = 0$ and $p_0^\mu \ep_\mu = 0$.
We can therefore write $q^\mu = q p_0^\mu$, etc.,
so that in what follows a quantity such as $q^2$ 
no longer denotes a square 
of the corresponding four-vector
but instead denotes the square 
of a scalar coefficient.
Various terms such as $\fslash{q}\ga^{\mu}\fslash{q}$ 
can then be shown to vanish upon contraction with $\ep_\mu$,
and after some algebra we find
that only terms quadratic in $p_0^\mu$ survive. 
These terms are finite, 
so the limit $d \rightarrow 4$ can be taken.
The result is:
\beq
A_1 + m^2 A_2 + m^4 A_3 &=& 
-\fr {i c_{\mu\nu} p_0^\mu p_0^\nu }{192 \pi^2 m^2} 
\Big[
\et^{\mu_1\mu_2} \et^{\mu_3\mu_4} 
( 80 q^2 - q (44p_1 + 32 p_2 + 12 p_3) + 4 p_1 p_{123} ) 
\nonumber \\ & & 
\hskip 60pt
+ \et^{\mu_1\mu_3} \et^{\mu_2\mu_4} 
( -112 q^2 + q (60 p_1 + 40 p_2 +  20 p_3) - 4 p_1 p_{123} )
\nonumber \\ & & 
\hskip 60pt
+ \et^{\mu_1\mu_4} \et^{\mu_2\mu_3} 
( 80 q^2 - q (52 p_1 + 32 p_2 + 20 p_3) + 4 p_1 p_{123} )
\Big],
\nonumber \\
B_1 + m^2 B_2 &=& 
\fr {i c_{\mu\nu} p_0^\mu p_0^\nu } {24\pi^2 m^2} 
\Big[
\et^{\mu_1\mu_2} \et^{\mu_3\mu_4} 
( 5 q^2 - q (6 p_1 + 3 p_2 + 3 p_3) +  p_1 p_{123} ) 
\nonumber \\ & &
\hskip 60pt
 + \et^{\mu_1\mu_3} \et^{\mu_2\mu_4} 
( -5 q^2 + q (6 p_1 + 3 p_2 +  3 p_3) -  p_1 p_{123} ) 
\nonumber \\ & & 
\hskip 60pt
+ \et^{\mu_1\mu_4} \et^{\mu_2\mu_3} 
( 3 q^2 - q (4 p_1 +  p_2 + 3 p_3) +  p_1 p_{123} )
\Big]. 
\label{ab}
\eeq
Inserting these expressions into Eq.\ \rf{AB},
we can explicitly integrate over the Feynman parameters
since these appear at most quadratically in the integrands. 
The result for $G= A+B$ is:
\beq
G^{\mu_1 \mu_2 \mu_3 \mu_4} 
&=& 
- \fr {i c_{\mu\nu} p_0^\mu p_0^\nu } {1440 \pi^2 m^2} 
\Big[
\et^{\mu_1\mu_2} \et^{\mu_3\mu_4} 
( 102 p_1^2 + 24 p_2^2 + 37 p_3^2  
+ 89 p_1 p_2 + 115 p_1 p_3 + 81 p_2 p_3 ) 
\nonumber \\ & &
\hskip 65pt
- \et^{\mu_1\mu_3} \et^{\mu_2\mu_4} 
( 126 p_1^2 + 48 p_2^2 + 41 p_3^2  
+ 133 p_1 p_2 + 119 p_1 p_3 + 93 p_2 p_3 ) 
\nonumber \\ & & 
\hskip 65pt
+ \et^{\mu_1\mu_4} \et^{\mu_2\mu_3} 
( 48 p_1^2 + 37 p_3^2  + 11 p_1 p_2 
+ 85 p_1 p_3 + 63 p_2 p_3 ) 
\Big].
\label{G}
\eeq

The full amplitude for the photon triple splitting
at this order is obtained by summing the contributions
from 24 graphs.
These consist of four diagrams of the form in Fig.\ 1a
with \c\ insertions on the different fermion propagators,
with three distinct topological channels for each diagram,
and two orientations of the fermion loop for each channel.
All these contributions can be derived from the result \rf{G} for $G$ 
by suitable index permutations. 
Defining $P$ as
$P( f(\mu_1, \mu_2, \mu_3, \mu_4, p_1, p_2, p_3, p_4) ) \equiv 
f(\mu_4, \mu_1, \mu_2, \mu_3, p_4, p_1, p_2, p_3)$, 
then the sum of the four \c\ insertions is given by 
$R = G + P(G) + P(P(G)) + P(P(P(G)))$ for one channel. 
To obtain different topological channels,
we interchange first $\mu_1$ and $\mu_2$ 
and subsequently $\mu_1$ and $\mu_4$ 
without changing the momentum indices. 
Hence,
the full contribution to the photon triple splitting amplitude $T_c$ is
$T_c = 2 ( R + C_{12}(R) + C_{14}(R))$,
where 
$C_{12}( f(\mu_1, \mu_2, \mu_3, \mu_4) ) 
= f(\mu_2, \mu_1, \mu_3, \mu_4)$,
$C_{14}( f(\mu_1, \mu_2, \mu_3, \mu_4) ) 
= f(\mu_4, \mu_2, \mu_3, \mu_1)$. 
Explicitly,
we find
\beq
T_c^{\mu_1 \mu_2 \mu_3 \mu_4} 
&=& 
- \fr {i c_{\mu\nu} p_0^\mu p_0^\nu } {60 \pi^2 m^2} 
\Big[
\et^{\mu_1\mu_2} \et^{\mu_3\mu_4} 
( 9  p_1^2  - 8 p_2^2 - p_3^2 
+ 2 p_1 p_2 + 16 p_1 p_3 - 18 p_2 p_3 ) 
\nonumber \\ & &
\hskip 60pt
+ \et^{\mu_1\mu_3} \et^{\mu_2\mu_4} 
( 4 p_1^2 - 8 p_2^2 + 4 p_3^2  
- 8 p_1 p_2 + 16 p_1 p_3 - 8 p_2 p_3 ) 
\nonumber \\ & & 
\hskip 60pt
+ \et^{\mu_1\mu_4} \et^{\mu_2\mu_3} 
( - p_1^2 - 8 p_2^2 + 9 p_3^2  - 18 p_1 p_2 
+ 16 p_1 p_3 + 2 p_2 p_3 ) 
\Big],
\label{tc}
\eeq
where we have used $p_1 + p_2 + p_3 + p_4 = 0$.
Thus, 
a nonzero coefficient \c\ yields a finite CPT-even contribution $T_c$
to the amplitude for photon triple splitting, 
even in the collinear limit. 

By a similar set of calculations we obtain  
a finite CPT-odd contribution $T_b$
to the amplitude for photon triple splitting 
from a nonzero coefficient \b, 
again in the collinear limit: 
\beq
T_b^{\mu_1 \mu_2 \mu_3 \mu_4} &=& 
\fr {b_\mu p_{0\nu}} {6 \pi^2 m^2} 
\Big[
(9 p_1 + 3 p_3) 
\et^{\mu_1\mu_2} \ep^{\mu_3\mu_4\mu \nu} 
- (5 p_1 + 4 p_2 + 3 p_3) 
\et^{\mu_1\mu_3} \ep^{\mu_2\mu_4\mu \nu} 
- (3 p_1 + p_3) 
\et^{\mu_2\mu_3} \ep^{\mu_1\mu_4\mu \nu} 
\Big].
\label{tb}
\eeq
\ntxt
Note that these results also imply nonzero contributions
in the off-shell, non-collinear case.
However,
the existence of nonzero amplitudes in the collinear limit
suffices to show that Schwinger's no-splitting result 
is vitiated in the presence of Lorentz violation 
\cite{fn2}.

The nonzero amplitudes \rf{tc}, \rf{tb} must arise 
from gauge-invariant terms in the effective action 
$S_{\rm eff}$ 
for the Lorentz-violating QED extension \rf{lag}. 
The form of $S_{\rm eff}$ is not known explicitly,
and it would be of interest to obtain it.
The amplitudes \rf{tc}, \rf{tb} are determined
in the collinear limit
and therefore cannot be used to deduce uniquely
the terms in $S_{\rm eff}$ from which they arise.
However,
$S_{\rm eff}$ could include expressions such as
$c_{\mu\nu} F^{\mu\ka} F_{\ka}^{\pt{\ka}\nu} 
\prt^{-2} F^{\al\be} F_{\al\be}$,
which for transverse photons in the collinear limit
generates terms of the structure necessary to reproduce 
the result \rf{tc}. 
Note that the appearance of the inverse box operator 
is to be expected in the effective action of a massless field
\cite{box}.

We close with brief remarks about an issue 
that is of some interest 
but lies beyond our present scope:
the determination of a physical rate for 
on-shell photon triple splitting.
For a massive particle,
the decay rate is defined in the rest frame 
with a kinematic factor inversely proportional to the mass,
so cannot trivially be extended to the massless case.
Since the theory \rf{lag} is observer Lorentz covariant,
it may be possible to define a physically consistent decay rate
in terms of the particle energy in the observer frame
\cite{ph}. 
There is also a separate kinematical issue to consider
because collinear momenta occupy a set of measure zero
in the phase space.  
However,
a physical analogy is provided by 
the collinear parametric down conversion 
of photons in optically active crystals,
for which photon multiple splitting 
has been experimentally observed 
\cite{gm}. 
The Lorentz violation
induces an analogous effective optical activity of the vacuum
\cite{ck,km}.
Equations \rf{tc} and \rf{tb} imply 
any nonzero decay rate would be quadratically suppressed
in the small coefficients $c_{\mu\nu}$ and $b_\mu$
\cite{jlm}.
Nonetheless,
Lorentz-violating photon degradation 
over cosmic distances 
might lead to observable effects on the redshift
\cite{fn3,pr},
conceivably even affecting open issues 
such as the origin of the cosmological constant. 

We thank R.\ Jackiw for discussions.
This work is supported in part
by the United States Department of Energy
under grant DE-FG02-91ER40661.

\end{multicols}

\begin{references}
\bibitem{fn1}
The methods adopted here can also be applied to verify directly 
the vanishing of the amplitude for photon triple splitting 
in conventional QED.
\bibitem{fn2}
O.W.\ Greenberg,
Phys.\ Rev.\ Lett.\ {\bf 89}, 231602 (2002)
has shown that CPT violation must be accompanied
by Lorentz violation,
so the latter can be viewed as the primary effect
even for $T_b$.
\bibitem{fn3}
For example, 
dimensional analysis 
suggests photons with $E\sim m$ have Hubble-scale lifetime 
if $c_{\mu\nu}$ is of order $10^{-19}$, 
an experimentally viable Planck-scale suppression.
\end{references}
\end{document}